\documentclass{rrparticle}
\usepackage{graphicx}
\graphicspath{ {./images/} }
\usepackage[rightcaption]{sidecap}
\usepackage{cite}
\usepackage{float}
\newcommand{\miktex}{\hbox{Mik\kern-.15em\TeX}}

\title{Two-body motion in central conservative force fields} 
\author[1]{O. Ianc}
\author[1]{A. Marin}
\author[1,a]{T. O. Cheche}
\affil[1]{University of Bucharest, Faculty of Physics, Măgurele, PO BOX MG11, 077125, Romania
\newline
Corresponding author email:$^a$ \em{tiberius.cheche@unibuc.ro}}

\newcommand*\diff{\mathop{}\!\mathrm{d}}
\DeclareMathAlphabet\mathbfcal{OMS}{cmsy}{b}{n}
\begin{document}
\maketitle
\begin{abstract}
The two-body motion for the gravitational and linear elastic central conservative force field is discussed by comparing the differential equations of motions. The inverse and direct square displacement dependency of the two force types is shown as emerging in the trajectories. The closed (for the gravitational field) and open (for the particular dumbbell elastic system considered) character of the trajectory is discussed.
\end{abstract}

\section{Introduction}
The relative motion and conservation laws are subjects present in all fields of physics. The calculus, both analytical and numerical, leads to conclusions that allow generalizations and explain the physics phenomena. C++, Fortran, Python programming languages, or Wofram Mathematica software (only to give a few examples) are valuable research tools which push the modeling to high accuracy limits. Mention of a few papers, some of them by one of the authors, could guide readers through such physics analyses involving general physics conservation laws and both analytical and numerical modeling. Thus, for example, the strain field in heterogeneous elastic structures is modeled by solving systems of differential equations by using Fortan and Mathematica programs 
\cite{cheche1, cheche2,cheche3}. Aerodynamical features of virtual wind turbines are studied by means of Comsol software\cite{voinea}. Optical dynamics in GaAs/AlAs and InAs/AlAs semiconductor quantum dots is simulated by using the conservation law of quantum angular momentum\cite{cheche4, cheche5,cheche6}. Image processing software is used 
for the analysis of data obtained in the optics physics laboratory \cite{radu}, 
The intrinsic spin Hall conductivity is considered by using the Hamiltonian of a two-dimensional electronic gas 
\cite{cheche7,cheche8}, 
and the electron transfer reaction rates and relaxation in dissipative systems are modeled by using the Nakajima–Zwanzig integral equations
\cite{cheche9,cheche10}. 
In pedagogical manner various motions in conservative fields are presented, such as the dynamics of a pulsejet engine in vertical motion in a uniform gravitational field without 
\cite{cheche11,cheche12,cheche13}.

In the present work the motion in conservative field is analyzed by comparing the two-body motion in gravitational \cite{saari, sanyal} and in a dumbbell elastic field\cite{dumbbell1, dumbbell2, dumbbell3}. For the gravitational motion we introduce the common two-body gravitational system (GTBS) and as an accompanying system, on which we focus our analysis, a dumbbell in which the two bodies are connected through a linear (massless and respecting Hooke’s law) elastic spring. We name this second system as elastic two-body system (ETBS). The mass of the two point masses is $m_{1,2}$ and the spring of initial undeformed length 2\emph{L} has the elastic constant \emph{k}. We discuss in more detail the ETBS motion and make a parallel between the equations of motion of GTBS and ETBS relative to the center of mass (CM) system.\
\
The structure of the article is as follows. In Section 2 the two-body motion is reviewed. Section 3 provides the equations for the numerical simulation. Then, in Section 4, the boundedness of the solutions is discussed. Section 5 concludes on the openness of the elastic orbits. Section 6 is for conclusions.

\section{Two-body motion}

As initial condition we consider body 1 as having the velocity $\mathbf{v}$, oriented as shown in the Figure 1.

\begin{figure}[h!tb]
\centering
\includegraphics[width=0.6\textwidth]{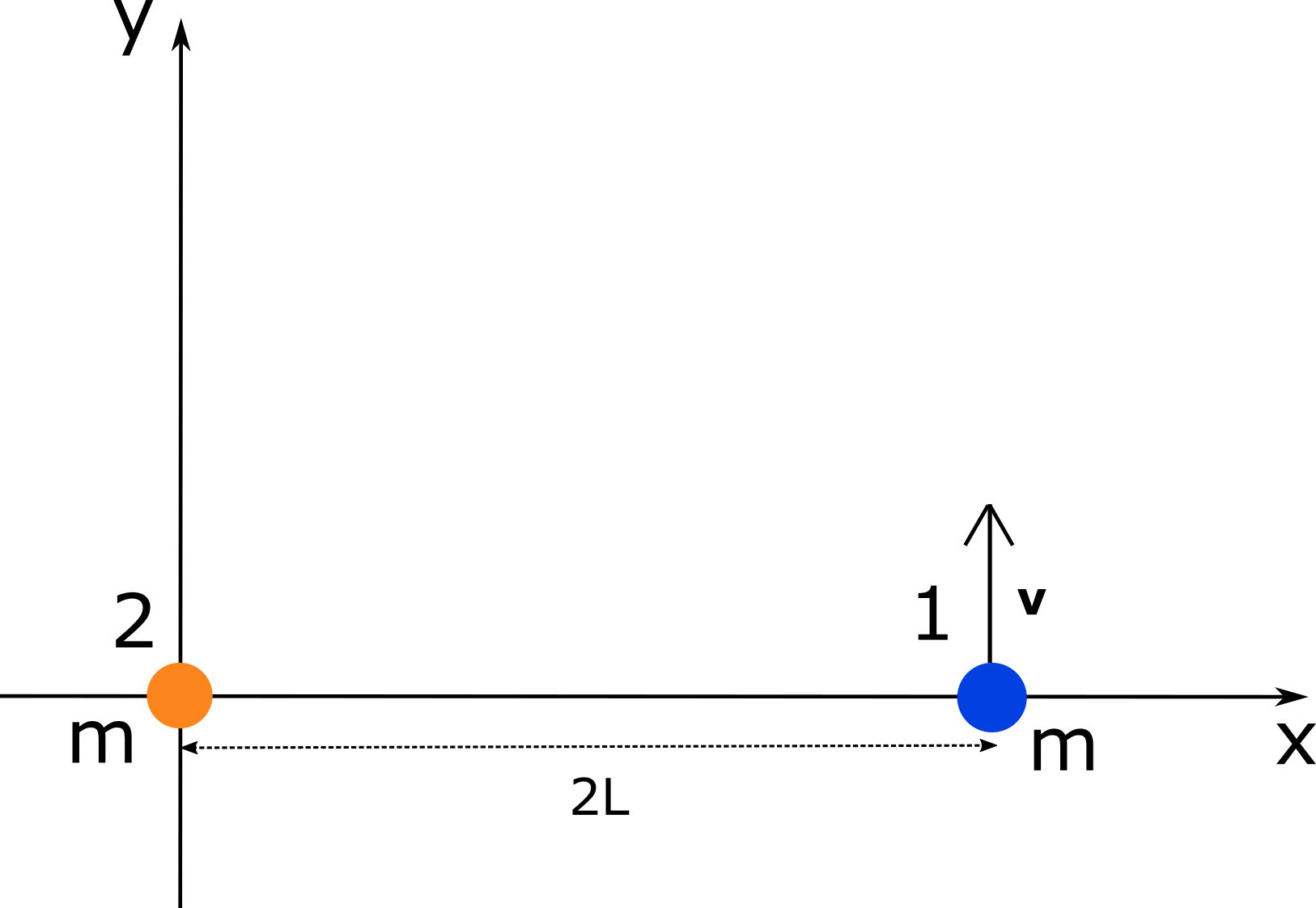}
\caption{Initial state of the systems}
\end{figure}

The linear momentum relative to the laboratory frame (LF) of the two-body system can be expressed by introducing the CM velocity $\mathbf{v_c}$ as follows:
    \begin{equation}
        m_1\mathbf{v}=(m_1+m_2)\mathbf{v_c}.
    \end{equation}
The angular momentum in LF is
\begin{equation}\begin{split}
\mathbf{J} &=\mathbf{r}_{1} \times \mathbf{p}_{1}+\mathbf{r}_{2} \times \mathbf{p}_{2}=\left(\mathbf{r}_{c}+\mathbf{r}_{1}^{\prime}\right) \times m_{1}\left(\dot{\mathbf{r}}_{c}+\dot{\mathbf{r}}_{1}^{\prime}\right)+\left(\mathbf{r}_{c}+\mathbf{r}_{2}^{\prime}\right) \times m_{2}\left(\dot{\mathbf{r}}_{c}+\dot{\mathbf{r}}_{2}^{\prime}\right) \\
&=\underbrace{\mathbf{r}_{c} \times\left(m_{1}+m_{2}\right) \dot{\mathbf{r}}_{c}}_{\mathbf{L_C}}+\underbrace{m_{1} \mathbf{r}_{1}^{\prime} \times \dot{\mathbf{r}}_{1}^{\prime}+m_{2} \mathbf{r}_{2}^{\prime} \times \dot{\mathbf{r}}_{2}^{\prime}}_{\mathbfcal{L}}=\mathbf{L}_{c}+\mu \mathbf{r} \times \mathbf{v}
.\end{split}\end{equation}
where we introduced $\mathbf{r_1}$, $\mathbf{r_2}$,  $\mathbf{p_1}$, $\mathbf{p_2}$ as the position vectors and momenta, respectively, relative to LF,
$\mu=\frac{m_1m_2}{m_1+m_2}$ (the reduced mass), $\mathbf{r}=\mathbf{r_1}-\mathbf{r_2}$, $\mathbf{r_c}=\frac{m_1\mathbf{r_1}+m_2\mathbf{r_2}}{m_1+m_2}$ (CM position vector relative to LF), $\mathbf{v}=\dot{\mathbf{r_1}}-\dot{\mathbf{r_2}}$.
Given the central force (which can exert no torque), the angular momenta are conserved:  
$\mathbfcal{\dot{L}}=\mathbf{\dot{L}_c}=\mathbf{\dot{J}}=0$.\\ 
At the initial time the (constant) angular momentum is
\begin{equation}
    \mathbf{J}=2Lm_1\mathbf{i}\times (v\mathbf{j})=2Lm_1v\mathbf{k}.
\end{equation}
With 
 $$\mathbf{r_{c}}(0)=\frac{m_12L\mathbf{i}}{m_1+m_2}$$
and eq. (1) we can write the (constant) angular momentum of CM, by considering the initial time, as
\begin{equation}
\mathbf{L_c}=\mathbf{r_{c}}(0)\times(m_1+m_2)\mathbf{v_c}=\frac{m_1^22Lv}{m_1+m_2}\mathbf{k}.
\end{equation}


Then the (constant) angular momentum of the two-body system with respect to the CM at the initial time is
\begin{equation}
    \mathbfcal{L}=\mathbf{J}-\mathbf{L_c}=2Lvm_1\left(1-\frac{m1}{m_1+m_2}\right)\mathbf{k}=\frac{2Lvm_1m_2}{m_1+m_2}\mathbf{k}=2\mu Lv\mathbf{k}
\end{equation}
and at a later time ($t>0$) by using the polar coordinates (with the unit vectors $\mathbf{e_r}$ and $\mathbf{e_\theta}$)
\begin{equation}\label{angmom}
  \mathbfcal{L}=\mu \mathbf{r}\times\mathbf{v}=\mu r\mathbf{e_r}\times(\dot r\mathbf{e_r}+r\dot\theta\mathbf{e_\theta})=\mu r^2\dot\theta\mathbf{k}.
\end{equation}
  The total energy in LF (written for the initial time) has the following expressions:
  \begin{equation}\label{energy}
      E_{tot}=
      \begin{cases}
    \frac{m_1v^2}{2}-\frac{\gamma m_1m_2}{2L} & \text{GTBS}; \\
    \frac{m_1v^2}{2}       & \text{ETBS}.
      \end{cases}
  \end{equation}
  One can rewrite eq. (\ref{energy}) by using its decomposition in the CM:
  \begin{equation}
      E_{tot}=\frac{(m_1+m_2)v_c^2}{2}+\underbrace{\frac{\mu\dot{r}^2}{2}+\frac{\mathcal{L}^2}{2\mu r^2}+U(r)}_E
  \end{equation}
  and
    \begin{align}\label{E}
      E&=E_{tot}-\frac{(m_1+m_2)v_c^2}{2}=
      \begin{cases}
        \frac{\mu v^2}{2}-\frac{\gamma m_1m_2}{2L} &\text{GTBS};\\
        \frac{\mu v^2}{2} &\text{ETBS}.
      \end{cases}
  \end{align}
  
  We can collect all terms depending on $r$ in a variable denoted as effective potential $U_{\mathrm{e}}$, and subsequently write:
  
  \begin{equation}\label{Ueff}
      E=\frac{\mu\dot{r}^2}{2}+\frac{\mathcal{L}^2}{2\mu r^2}+U(r)=\frac{\mu\dot{r}^2}{2}+U_\mathrm{e}(r)
  \end{equation}
One can now solve eq. (\ref{Ueff}) for $\dot{r}$ and obtain:
\begin{equation}\label{rdot}
    \dot{r}=\pm\sqrt{\frac{2}{\mu}\left[E-U(r)-\frac{\mathcal{L}^2}{2\mu r^2}\right]}.
\end{equation}

The next table summarizes the differences between the two scenarios of interest, where we denoted $\alpha\equiv\gamma m_1m_2$:\\
  \begin{table}[h!t]%
\centering
\begin{tabular}{|c|c|c|}
\hline
{} & $E$ & $U(r)$ \cr
\hline
GTBS &$\frac{m_1v^2}{2}-\frac{\alpha}{2L}$ & $-\frac{\alpha}{r}$ \cr
\hline
ETBS &$\frac{m_1v^2}{2}$ & $\frac{k(r-2L)^2}{2}$ \cr
\hline
\end{tabular}
\label{comparison}
\caption{Comparison between gravitational and elastic interaction in the two-body problem.}
\end{table}

\section{Boundedness of the solutions}
Equation (\ref{rdot}) allows us to impose the condition for the existence of solutions:
\begin{equation}\label{exist}
    \dot{r}\in\mathbb{R} \iff E\ge U_\mathrm{e}(r)=U(r)+\frac{\mathcal{L}^2}{2\mu r^2}.
\end{equation}
For GTBS we have
\begin{equation}\label{Ueffg}
    U_\mathrm{e}(r)=-\frac{\alpha}{r}+\frac{\mathcal{L}^2}{2\mu r^2}
\end{equation}
and this function vanishes for $r=\frac{\mathcal{L}^2}{2\mu\alpha}$. By setting the derivative of eq. (\ref{Ueffg})
\begin{equation}
    U'_\mathrm{e}(r)=\frac{\alpha}{r^2}-\frac{\mathcal{L}^2}{\mu r^3}.
\end{equation}
to zero one obtains an extremum point for $U_\mathrm{e}(r)$ at $r=\frac{\mathcal{L}^2}{\mu\alpha}$. To establish its nature, this value must be plugged into the expression for the second derivative of $U_\mathrm{e}(r)$:
\begin{equation}
    U''_\mathrm{e}(r)=-\frac{2\alpha}{r^3}+\frac{3\mathcal{L}^2}{\mu r^4}.
\end{equation}
Since $U''_\mathrm{e}(\frac{\mathcal{L}^2}{\mu\alpha})=\frac{\alpha^4\mu^3}{L^6}>0$, the extremum is a minimum point and the orbit is stable. Therefore, $r_{min}=\frac{\mathcal{L}^2}{\mu\alpha}$. 
The analysis in Table \ref{t_Ueffg} shows that the system is in a bound state for $0>E>\frac{-\mu\alpha^2}{2\mathcal{L}^2}$.\\

\begin{table}[h!t]%
\centering
\begin{tabular}{|c|c c c c c c c|}
\hline
r & 0 & & $\frac{\mathcal{L}^2}{2\mu\alpha}$& &$\frac{\mathcal{L}^2}{\mu\alpha}$& &$+\infty$
\cr
\hline
$U_\mathrm{eff}(r)$ & $+\infty$ &$+$&$0$&$-$ &$\frac{-\mu\alpha^2}{2\mathcal{L}^2}$ &$-$&$0$\cr
\hline
 $U'_\mathrm{eff}$& $-$& &$-$      &     &$0$  & &   $0^+$\cr
\hline
 $U_\mathrm{eff}$& & &$\searrow$&  &min &$\nearrow$   &$\rightarrow$\cr
\hline
\end{tabular}
\label{t_Ueffg}
\caption{Study of $U_\mathrm{eff}(r)$ for the gravitational field.}
\end{table}
Next, we analyze the effective potential for ETBS.
\begin{subequations}
\begin{align}
      U_\mathrm{e}(r)=\frac{k(r-2L)^2}{2}+\frac{\mathcal{L}^2}{2\mu r^2} (>0\text{, }\forall r\in\mathbb R).\\
    U_\mathrm{e}(2L)=\frac{\mathcal{L}^2}{2\mu (2L)^2}=\frac{4\mu^2L^2v^2}{2\mu 4L^2}=\frac{\mu v^2}{2}=E.\\
    U_\mathrm{eff}\mid_{r\to+\infty}=+\infty\text{; } U_\mathrm{eff}\mid_{r\to 0^+}=+\infty.
\end{align}
\end{subequations}
\begin{subequations}
\begin{align}
    U'_\mathrm{e}(r)=k(r-2L)-\frac{\mathcal{L}^2}{\mu r^3}.\\
    U'_\mathrm{e}(2L)=-\frac{\mathcal{L}^2}{\mu (2L)^3}<0.
\end{align}
\end{subequations}
The compatibility condition imposes for this case as $E\ge U_\mathrm{e}\text{, }\forall r\in[2L,r_{max}]$ and the trajectory is bounded and stable (the second derivative is positive); $r_{max}>2L$ is the second root of $U_\mathrm{e}(r)=E$.

\begin{table}[h!t]%
\centering
\begin{tabular}{|c|c c c c c c c|}
\hline
r & 0 & & $2L$ & & $r_{max}$ & &$+\infty$
\cr
\hline
$U_\mathrm{e}(r)$ & $+\infty$ & + & $E$ & + & $E$ & + &$+\infty$\cr
\hline
 $U'_\mathrm{e}$& $-\infty$ & & $-\frac{\mathcal{L}^2}{8 \mu L^3}$      &     $0$  & + &  & $+\infty$\cr
\hline
 $U_\mathrm{e}$& & & $\searrow$  & $\mathrm{min}$ & $\nearrow$ &    &\cr
\hline
\end{tabular}
\label{t_Ueffe}
\caption{Study of $U_\mathrm{e}(r)$ for the elastic field.}
\end{table}

\section{On the shape and closedness of the elastic orbits}
Bertrand's theorem \cite{bertrand, bertrand1, bertrand2} states that in a central field of potential energy $U(r)$, all bounded orbits are closed if and only if the potential energy can be written as $kr^2$ or $-\frac{k}{r}$ for $k>0$. For ETBS the elastic potential energy, discarding an arbitrary constant, can be written as $U (r) = kr^2/2 -2kr$, so it does not satisfy the Bertrand’s
theorem conditions. Consequently, while GTBS has bounded and closed orbits ETBS shows bounded but open trajectories. As the minimum $\leftrightarrow$ maximum distance between the bodies of ETBS is $2L\leftrightarrow r_{max}$, the trajectories with respect to the CM are bounded by concentric circles of radii $2m_2L/(m_1+m_2)$ and $m_2r_{max}/(m_1+m_2)$ for the body of $m_1$ and similar for the body of mass $m_2$.\\
Going into deeper details, the proof of Bertrand theorem uses as a starting point a lemma\cite{analytmec} which establishes what is the possible form of the potential when in a central field all orbits, close to a circular orbit, are closed. Thus, it is known the following formula between the angle of a pericenter (minimum of $r$) and an apocenter (maximum of $r$):
\begin{equation}
    \Phi=\int\limits_{r_{min}}^{r_{max}}\frac{\mathcal{L}}{r^2\sqrt{2\mu[E-U_{\mathrm{e}}(r)]}}\diff r
\end{equation}
If one denotes $r_c$ as the radius of a possible circular orbit (i.e. $U'_\mathrm{e}(r_c)=0$), for an orbit not far to a circular orbit of radius $r_c$, to a first approximation
\begin{equation}
    \Phi=\pi \sqrt{\frac{U'(r_c)}{r_c U''(r_c)+3 U'(r_c)}}
\end{equation}
To establish the form of potential for closed orbits near a circular orbit one imposes that $\Phi$ is independent of $\mathcal{L}$ or equivalently of $r_c$. Thus for $\Phi$ to be a constant it should be independent of $r_c$ and real. Therefore one concludes that 
\begin{equation}
    \frac{U'(r_c)}{r_c U''(r_c)+3 U'(r_c)}=c \ge 0, \text{ constant}
\end{equation}
Substitution of $U(r)$ by the expression from Table 1 gives
    \begin{align}\label{c}
      c&=
      \begin{cases}
        1 &\text{GTBS};\\
        \frac{r_c-L}{4 r_c - 3 L} &\text{ETBS}.
      \end{cases}
  \end{align}
As $c$ is not a constant (unless not physical case $L=0$) for the ETBS the mentioned lemma can not be used in the proof of Bertrand theorem: the ETBS potential can not generate close trajectory even for trajectories close to circular orbits.

\section{Numerical simulations of the two systems}


Our goal is to represent the trajectory of the two-body system relative to the CM. This can be achieved by separating the variables in eq. (\ref{angmom}):
\begin{equation}\label{rdot2}
    \dot{r}=\frac{\diff r}{\diff \theta}\dot\theta=\frac{\diff r}{\diff \theta}\frac{\mathcal{L}}{\mu r^2}.
\end{equation}
By integration one can obtain $t(r)$.
For the gravitational field the integral can be expressed in terms of elementary functions \cite{goldstein, arnold}, but inverting it in order to obtain $r(t)$ is difficult to do analytically. A numerical solution of eq. (\ref{rdot}) can be considered instead, but the branched form of the equation requires a careful choice of the integration method and the existence of stable points where $\dot{r}=0$, which require a jump in order for the integral to converge, deem as unfeasible this way of numerically simulating the systems.\\
Therefore, we resorted to using the Newton's second law along the radial axis (the projection on the angular axis leads to $\mathbfcal{L}$ conservation) which can be obtained by either  using the Euler-Lagrange formalism or directly the radial acceleration expression: 

\begin{equation}\label{numsol1}
-\frac{dU(r)}{dr}=\mu (\ddot{r}-r \dot{\theta}^2)=\mu (\ddot{r}-\frac{\mathbfcal{L}^2}{2\mu r^3})
\end{equation}
that is

\begin{equation}\label{numsol2}
    \mu \ddot{r} = \frac{\mathbfcal{L}^2}{\mu r^3} - \frac{dU(r)}{dr}=-\frac{dU_e(r)}{dr}.
\end{equation}
With given expressions for $U(r)$ and $\mathbfcal{L}$  the first equality in eq. (24) is integrated numerically for each of the two cases considered. To plot the trajectory as a function of time we also integrate numerically the last equality in eq. (6) to obtain $\theta(t)$ and then transform back from polar to Cartesian coordinates by
    $x(t) = r(t)\cos(\theta(t)),  y(t) = r(t)\sin(\theta(t))$.\\
For the integration we use the Wolfram Mathematica software.
\begin{figure}[H] 
\centering
\includegraphics[width=0.6\textwidth]{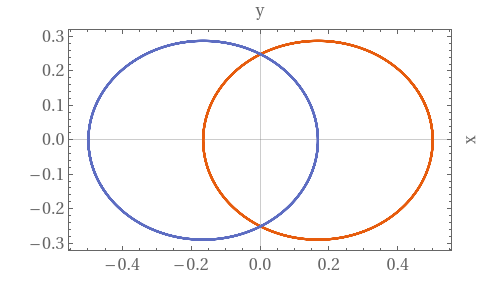}
\caption{Example solution for the gravitational case (GTBS) - Keplerian orbits (SI units):\\ $m_1=m_2=1$, $L=1$, $k=1$, $v=0.1$}
\end{figure}
In Figure 2, an orbit is presented for GTBS (animation available \href{https://github.com/octavianc27/twobodysimulations/blob/main/gravitational.gif}{here}). In Figure 3, bounded open trajectory is shown for ETBS (animation available \href{https://github.com/octavianc27/twobodysimulations/blob/main/elastic.gif}{here}). The code we used can be found at https://github.com/octavianc27/twobodysimulations.

\begin{figure}[H] 
\centering
\includegraphics[width=0.6\textwidth]{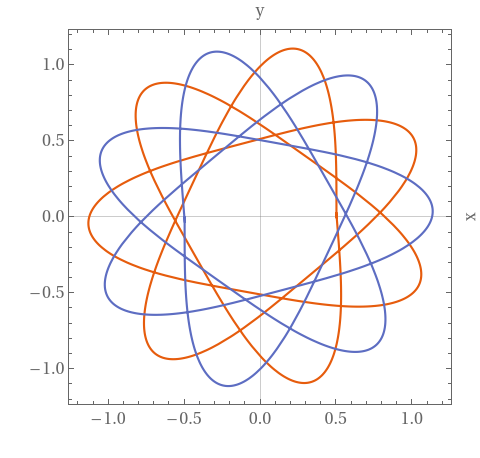}
\caption{Example solution for the elastic case (GTBS) - open orbits (SI units):\\ $m_1=m_2=1$, $L=1$, $k=1$, $v=1$}
\end{figure}
\section{Conclusion}
There are similarities between the motions in gravitational and elastic field. Thus, according to Bertrand's theorem, GTBS can have bounded and closed orbits if Kepler's condition of motion are fulfilled and, in this case, the trajectory of each body relative to CM is an ellipse. Similarly, according to the same Bertrand's theorem, the trajectory of one-body moving in linear elastic force field is bounded and close. On another hand, differences arise when the elastic system is the ETBS we discussed. In this last case, we obtain the trajectory is bounded but not closed, and this is the novelty of the work. We described the motion relative to the CM of the two-body system. Technically, we found easier to solve numerically the second order differential equation for the radial motion than the first order differential equation obtained from the energy conservation for $r(t)$.


\end{document}